\documentclass[10pt]{article}

\usepackage{graphicx}%
\usepackage{amsmath,amssymb,amsthm,upref,bm}%
\usepackage{labelfig}%
\usepackage{mathrsfs}

\DeclareMathAlphabet{\varmathbb}{U}{pxsyb}{m}{n}

\newtheorem{proposition}{Proposition}%
\newtheorem{definition}{Definition}%

\newcommand{\MF}[1]{\mathop{#1\vrule height0.45ex width0pt}\nolimits}

\newcommand{\pd}[3][]{\mathchoice{\raise-0.5pt\hbox{$\partial$}%
\vphantom{\partial}_{\mkern-1.5mu#2}^{\mkern0.4mu#1}\mkern0.3mu}%
{\raise-0.5pt\hbox{$\partial$}%
\vphantom{\partial}_{\mkern-1.5mu#2}^{\mkern0.4mu#1}\mkern0.3mu}%
{\raise-0.5pt\hbox{$\scriptstyle\partial$}%
\vphantom{\partial}_{\mkern-1.7mu#2}^{\mkern0.1mu#1}\mkern0.1mu}%
{\raise-0.5pt\hbox{$\scriptscriptstyle\partial$}%
\vphantom{\partial}_{\mkern-1.7mu#2}^{\mkern0.1mu#1}\mkern0.1mu}#3}
 
\newcommand{\D}{\mathrm{d}\kern0.2pt}%
\newcommand{\ii}{\kern0.05em\mathrm{i}\kern0.05em}% 
\newcommand{\E}[1]{\textrm{e}^{#1}}%
\renewcommand{\vec}[1]{\bm{#1}}%
\newcommand{\RR}{\varmathbb{R}}%
\newcommand{\Cb}{\varmathbb{C}}%

\newcommand\mhat[1]{\widehat{#1}}

\def\transp{\mathsf{T}}

\begin{document}

\baselineskip=4.4mm

\makeatletter

\title{\bf Two-dimensional water waves \\ in the presence of a freely floating body:
conditions for the absence \\ of trapped modes}

\author{Nikolay Kuznetsov}

\date{}

\maketitle

\vspace{-10mm}

\begin{center}
Laboratory for Mathematical Modelling of Wave Phenomena, \\ Institute for Problems
in Mechanical Engineering, Russian Academy of Sciences, \\ V.O., Bol'shoy pr. 61,
St. Petersburg 199178, Russian Federation \\ E-mail: nikolay.g.kuznetsov@gmail.com
\end{center}

\begin{abstract}
The coupled motion is investigated for a mechanical system consisting of water and a
body freely floating in it. Water occupies either a half-space or a layer of
constant depth into which an infinitely long surface-piercing cylinder is immersed,
thus allowing us to study two-dimensional modes. Under the assumption that the
motion is of small amplitude near equilibrium, a linear setting is applicable and
for the time-harmonic oscillations it reduces to a spectral problem with the
frequency of oscillations as the spectral parameter. It is essential that one of the
problem's relations is linear with respect to the parameter, whereas two others are
quadratic with respect to it.

Within this framework, it is shown that the total energy of the water motion is
finite and the equipartition of energy holds for the whole system. On this basis, it
is proved that no wave modes can be trapped provided their frequencies exceed a
bound depending on cylinder's properties, whereas its geometry is subject to some
restrictions and, in some cases, certain restrictions are imposed on the type of
mode.
\end{abstract}

\setcounter{equation}{0}

\section{Introduction}

This paper continues the rigorous study (initiated in \cite{NGK10}) of the coupled
time-harmonic motion of the mechanical system which consists of water and a rigid
body freely floating in it. The former is bounded from above by a free surface,
whereas the latter is assumed to be an infinitely long cylinder which allows us to
investigate two-dimensional modes orthogonal to its generators. The body is
surface-piercing and no external forces acts on it (for example, due to constraints
on its motion). The water domain is either infinitely deep or has a constant finite
depth; the surface tension is neglected on the free surface of water whose motion is
irrotational. The motion of the whole system is supposed to be small-amplitude near
equilibrium which allows us to use a linear model.

In the framework of the linear theory of water waves, the time-dependent problem
describing the coupled motion of water and a freely floating surface-piercing rigid
body was developed by \cite{John1}. However, his formulation was rather cumbersome,
and so during the second half of the 20th century the main efforts were devoted to
various problems involving fixed bodies instead of freely floating ones (see the
summarising monograph by \cite{LWW}). The cornerstone was laid by \cite{John2}
himself who proved the first result guaranteeing the absence of trapped modes at all
frequencies provided an immersed obstacle has a fixed position and is subject to a
geometric restriction now usually referred to as John's condition. In the
two-dimensional case, it includes the following two requirements: (i) there is only
one surface-piercing cylinder in the set of cylinders forming the obstacle; (ii) the
whole obstacle is confined within the strip between two vertical lines through the
points, where the surface-piercing contour intersects the free surface of water, the
part of bottom (when the depth is finite) is horizontal outside of this strip.

\cite{SU} demonstrated that if condition (i) holds, then condition (ii) can be
replaced by a weaker one. Namely, if the depth is infinite, then the whole obstacle
must be confined to the angular domain between the lines inclined at $\pi/4$ to the
vertical and going through the two points, where the surface-piercing contour
intersects the free surface. If the depth is finite, then it is required that the
whole obstacle is confined to a smaller angular domain between the lines going
through the same two points, but inclined at a certain angle to the vertical that is
a little bit less than $\pi/4$. The results of \cite{SU} and \cite{John2} are
illustrated in \cite{LWW}; see pp.~125, 126 and 137, respectively.

In \cite{K}, another geometric condition alternative to (ii) was found which
together with (i) guarantees the absence of trapped modes at all frequencies for
fixed bodies. This condition does not impose any restriction on the angle between
the surface-piercing contour and the free surface (arbitrarily small angles are
admissible), but this is achieved at the expense that the wetted contour is subject
to a certain point-wise restriction (it must be transversal to curves \eqref{eq:ca}
in a certain definite fashion).

On the other hand, condition (i) is essential for the absence of trapped modes. This
became clear when \cite{MM} constructed an example of such a mode for which purpose
she applied the so-called semi-inverse method (see, for example, \cite{KM1} for its
brief description). Her example involves two fixed surface-piercing cylinders each
of which satisfies the modified condition (ii) of \cite{SU}, but they are separated
by a nonzero spacing. Another example of a mode trapped by two fixed
surface-piercing cylinders was found by \cite{MK}. Subsequently, \cite{NGK10} proved
that the latter cylinders can be considered as two immersed parts of a single body
which freely floats in trapped waves, but remains motionless.

\begin{figure}
\begin{center}
\vspace{-1mm}
\SetLabels
\L (0.525*0.94) $y$ \\
\L (0.93*0.8) $x$ \\
\L (0.8*0.8) $F$ \\
\L (0.615*0.8) $+a$ \\
\L (0.312*0.8) $-a$ \\
\L (0.15*0.8) $F$ \\
\L (0.56*0.54) $B$ \\
\L (0.32*0.44) $S$ \\
\endSetLabels
\leavevmode
\strut\AffixLabels{\includegraphics[width=65mm]{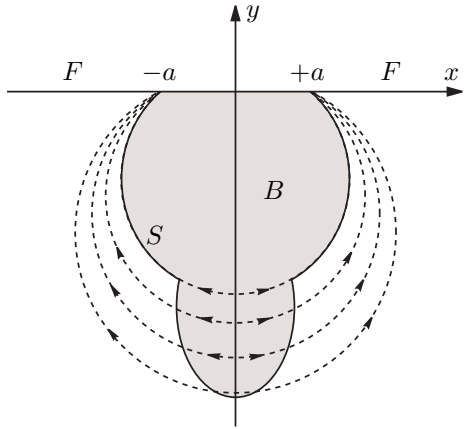}}
\end{center}
\vspace{-6mm} \caption{A definition sketch of the submerged cross-section of a
cylinder.} \vspace{-1mm}
\label{fig1}
\end{figure}

During the past decade, the problem of the coupled time-harmonic motion of water and
a freely floating rigid body has attracted much attention. Along with the just
mentioned paper \cite{NGK10}, rigorous results were obtained in \cite{KM1}, where a
brief review of related papers is given. However, the substantial part of work
concerns the study of trapped modes and the corresponding trapping bodies and only
the paper \cite{KM} has been focussed on conditions eliminating trapped modes in
the case when a surface-piercing or totally submerged body is present (for a
surface-piercing body the original proof of \cite{John2} was essentially
simplified). In the present paper, our aim is to fill in this gap at least
partially.

In the present note, we find conditions on the frequency so that they guarantee that
no modes (or some specific modes) are trapped by a freely floating body provided its
geometry satisfies the assumptions used in \cite{SU} and \cite{K} for establishing
the absence of modes trapped by the same body being fixed.

\vspace{-6mm}

\section{Statement of the problem}

Let the Cartesian coordinate system $(x,y)$ in a plane orthogonal to the generators
of a freely floating infinitely long cylinder be chosen so that the $y$-axis is
directed upwards, whereas the mean free surface of water intersect this plane along
the $x$-axis, and so the cross-section $W$ of the water domain is a subset of
$\RR^2_- = \{ x \in \RR, \, y<0 \}$. Let $\mhat{B}$ denote the bounded
two-dimensional domain whose closure is the cross-section a floating cylinder in its
equilibrium position. Let both the immersed part $B = \mhat{B} \cap \RR^2_-$ and the
above-water part $\mhat{B} \setminus \overline{\RR^2_-}$ be nonempty domains and $D
= \mhat{B} \cap \partial \RR^2_-$ be a nonempty interval of the $x$-axis, say $\{ x
\in (-a, a), \, y=0 \}$ (see figure~\ref{fig1}). We suppose that $W$ is either
$\RR^2_- \setminus \skew2\overline{B}$ when water has infinite depth (see
figure~\ref{fig1}) or $\{ x \in \RR, \, -h<y<0 \} \setminus \skew2\overline{B}$,
where $h > b_0 = \sup_{(x,y) \in B} |y|$, when water has constant finite depth. We
suppose that $W$ is a Lipschitz domain, and so the unit normal $\vec{n}$ pointing to
the exterior of $W$ is defined almost everywhere on $\partial W$. Finally, by $S =
\partial \mhat{B} \cap \RR^2_-$ and $F = \partial \RR^2_- \setminus \skew2 \overline
{D}$ we denote the wetted contour and the free surface at rest, respectively; if
water has finite depth, then $H = \{ x \in \RR , \, y = - h \}$ is the bottom's
cross-section.

For describing the small-amplitude coupled motion of the system it is standard to
apply the linear setting in which case the following first-order unknowns are used.
The velocity potential $\MF{\Phi} (x,y;t)$ and the vector-column $\MF{\vec{q}} (t)$
describing the motion of body whose three components are as follows:

\vspace{1mm}

\noindent $\bullet$ $q_1$ and $q_2$ are the displacements of the centre of mass in
the horizontal and vertical directions, respectively, from its rest position $\bigl(
x^{(0)}, y^{(0)} \bigr)$;

\noindent $\bullet$ $q_3$ is the angle of rotation about the axis that goes through
the centre of mass orthogonally to the $(x,y)$-plane (the angle is measured from the
$x$- to $y$-axis).

\vspace{1mm}

\noindent We omit relations governing the time-dependent behaviour (see details in
\cite{NGK10}), and turn directly to the time-harmonic oscillations of the system for
which purpose we use the ansatz 
\begin{equation}
 \bigl( \MF{\Phi} (\vec{x},y,t), \vec{q}(t) \bigr) = {\rm Re} \bigl\{ \E{-\ii\omega t} 
 \bigl( \MF{\varphi} (\vec{x},y), \ii \vec{\chi} \bigr) \bigr\} ,
 \label{eq:ansatz}
\end{equation}
where $\omega > 0$ is the radian frequency, $\varphi \in \MF{H^1_{\mathrm{loc}}}(W)$
is a complex-valued function and $\vec{\chi} \in \Cb^3$. To be specific, we first
assume that $W$ is infinitely deep in which case the problem for $\bigl( \varphi,
\vec{\chi} \bigr)$ is as follows:\vskip-6mm
\begin{gather}
 \nabla^2 \varphi = 0 \quad \mbox{in} \ W ,
 \label{eq:1}\\
 \pd{y} \varphi - \nu \varphi = 0 \quad \mbox{on} \ F, \quad \mbox{where} \ 
 \nu = \omega^2 / g ,
 \label{eq:2}\\
 \pd{\vec{n}} \varphi = \omega \, \vec{N}^\transp \vec{\chi} \ \Big(\!\! = \omega 
 \sum_1^3 N_j \chi_j \Big) \quad \mbox{on} \ S , 
 \label{eq:4}\\
 \nabla \varphi \to 0 \quad \mbox{as} \ y \to -\infty ,
 \label{eq:3}\\
 \int_{W\cap\{|x|=b\}} \bigl| \pd{|x|} \varphi - \ii \nu \varphi \bigr|^2\,
 \D{}s = \MF{o}(1) \quad \mbox{as} \ b \to \infty ,
 \label{eq:6}\\
 \omega^2 \bm{E} \vec{\chi} = - \omega \int_{S} \varphi \vec{N} \,
 \D{}s + g \, \bm{K} \vec{\chi} . \label{eq:5}
\end{gather}
Here $\nabla = (\pd{x}, \pd{y})$ is the spatial gradient, $g > 0$ is the
acceleration due to gravity that acts in the direction opposite to the $y$-axis;
$\vec{N} = (N_1, N_2, N_3)^\transp$ (the operation $^\transp$ transforms a
vector-row into a vector-column and vice versa), where $(N_1, N_2)^\transp =
\vec{n}$, $N_3 = \left( x - x^{(0)}, y - y^{(0)} \right) \times \vec{n}$ and
$\times$ stands for the vector product. In the equations of body's motion
\eqref{eq:5}, the $3\!\times\!3$ matrices are as follows:
\begin{equation}
 \MF{\bm{E}} =
\begin{pmatrix}
 I^M & 0 & 0 \\
 0 & I^M & 0 \\
 0 & 0 & I^M_2
\end{pmatrix} \quad {\rm and} \quad \MF{\bm{K}} =
\begin{pmatrix}
 0 & 0 & 0 \\
 0 & I^D & I^D_x \\
 0 & I^D_x & I^D_{xx} + I^S_y
\end{pmatrix} .
\label{eq:EK}
\end{equation}
The positive elements of the mass/inertia matrix $\MF{\bm{E}}$ are
\[ I^M = \rho_0^{-1} \int_{\mhat{B}} \MF{\rho}(x,y) \, \D x \D y \ \ \mbox{and} \ \
I^M_2 = \rho_0^{-1} \int_{\mhat{B}} \MF{\rho}(x,y) \Big[ \left( x - x^{(0)}
\right)^2 + \left( y - y^{(0)} \right)^2 \Big] \D x \D y ,
\]
where $\MF{\rho}(x,y) \geq 0$ is the density distribution within the body and
$\rho_0 > 0$ is the constant density of water. In the right-hand side of relation
\eqref{eq:5}, we have forces and their moments. In particular, the first term is due
to the hydrodynamic pressure, whereas the second one is related to the buoyancy
(see, for example, \cite{John1}); the non-zero elements of the matrix $\bm{K}$
are
\begin{gather*}
I^D = \int_D \D x > 0, \quad I^D_x = \int_D \big( x - x^{(0)} \big) \D x , \\
I^D_{xx} = \int_D \big( x - x^{(0)} \big)^2 \D x > 0 , \quad I^S_y = \int_S \big( y
- y^{(0)} \big) \D x \, \D y .
\end{gather*}
Note that the matrix $\bm{K}$ is symmetric.

In relations \eqref{eq:2}, \eqref{eq:4} and \eqref{eq:5}, $\omega$ is a spectral
parameter which is sought together with the eigenvector $(\varphi,\vec{\chi})$.
Since $W$ is a Lipschitz domain and $\varphi \in \MF{H^1_{\mathrm{loc}}}(W)$,
relations \eqref{eq:1}--\eqref{eq:4} are, as usual, understood in the sense of the
following integral identity:
\begin{equation}
 \int_{W} \nabla \varphi \nabla \psi \,\D x \D{}y = \nu 
 \int_{F} \varphi \, \psi \, \D x + \omega \int_{S} \psi \, 
 \vec{N}^\transp \vec{\chi} \, \D{}s ,
\label{eq:intid}
\end{equation}
which must hold for an arbitrary smooth $\psi$ having a compact support in
$\overline W$. Finally, relations \eqref{eq:3} and \eqref{eq:6} specify the
behaviour of $\varphi$ at infinity. The first of these means that the velocity field
decays with depth, whereas the second one yields that the potential given by formula
\eqref{eq:ansatz} describes outgoing waves. This radiation condition is the same as
in the water-wave problem for a fixed obstacle (see, for example, \cite{John2}).

The relations listed above must be augmented by the following conditions concerning the
equilibrium position:

\vspace{1mm}

\noindent $\bullet$ The mass of the displaced liquid is equal to that of the body:
$I^M = \int_B \D x \D{}y$ (Archimedes' law); $\bullet$ The centre of buoyancy lies
on the same vertical line as the centre of mass: $\int_B \bigl(x - x^{(0)} \bigr) \,
\D x \D{}y = 0$; $\bullet$ The matrix $\bm{K}$ is positive semi-definite; moreover,
the $2 \times 2$ matrix $\bm{K}'$ that stands in the lower right corner of $\bm{K}$
is positive definite (see \cite{John1}).

\vspace{1mm}

\noindent The last of these requirements yields the stability of the body's
equilibrium position, which follows from the results formulated, for example, by
\cite{John1}, \S\,2.4. The stability is understood in the classical sense that an
instantaneous, infinitesimal disturbance causes the position changes which remain
infinitesimal, except for purely horizontal drift, for all subsequent times.

In conclusion of this section, we note that relations \eqref{eq:3} and \eqref{eq:6}
must be amended in the case when $W$ has finite depth. Namely, the no flow condition
\begin{equation}
\pd{y} \varphi = 0 \quad \mbox{on} \ H \label{eq:H}
\end{equation}
replaces \eqref{eq:3}, whereas $\nu$ must be changed to $k_0$ in \eqref{eq:6}, where
$k_0$ is the unique positive root of $k_0 \MF{\tanh}(k_0 h) = \nu$.

\section{Equipartition of energy, trapped modes and conditions guaranteeing their
absence}

\subsection{Equipartition of energy}

It is known (see, for example, \cite[\S\,2.2.1]{LWW}), that a potential,
satisfying relations \eqref{eq:1}, \eqref{eq:2}, \eqref{eq:3} and \eqref{eq:6}, has
the asymptotic representation at infinity of the same type as Green's function.
Namely, if $W$ has infinite depth, then
\begin{equation}
 \MF{\varphi} (x, y) = \MF{A}_\pm (y) \, \E{\ii \nu |x|} + 
 \MF{r}_\pm (x, y) , \ \mbox{where} \ |r_\pm|^2, \, |\nabla r_\pm| = \MF{O} \bigl(
 [x^2 + y^2]^{-1} \bigr) \ \mbox{as} \ x^2 + y^2 \to \infty ,
\label{eq:uas}
\end{equation}
and the following equality holds
\begin{equation}
 \nu \int_{-\infty}^0 \left( |\MF{A}_+ (y)|^2 + |\MF{A}_- (y)|^2 \right) \D{} y =
 - {\rm Im} \int_S \overline{\varphi}\,\pd{\vec{n}} \varphi \, \D{}s .
\label{eq:ener}
\end{equation}

Assuming that $\bigl( \MF{\varphi} , \vec{\chi} \bigr)$ is a solution of problem
\eqref{eq:1}--\eqref{eq:5}, we rearrange the last formula using the coupling
conditions \eqref{eq:4} and \eqref{eq:5}. First, transposing the complex conjugate of
equation \eqref{eq:5}, we get
\[\omega^2 \left( \bm{E} \overline{\vec{\chi}} \right)^\transp = - \omega
 \int_{S} \overline{\varphi} \vec{N}^\transp \, \D{}s + g \left( \bm{K}
 \overline{\vec{\chi}} \right)^\transp .
\]
This relation and condition \eqref{eq:4} yield that the inner product of both sides with
$\vec{\chi}$ can be written in the form:
\begin{equation}
 \omega^2 \, \overline{\vec{\chi}}^\transp \bm{E} \vec{\chi} - 
 g \overline{\vec{\chi}}^\transp 
 \bm{K} \vec{\chi} = - \int_{S} \overline{\varphi} \, \pd{\vec{n}} \varphi \,\D{}s .
\label{eq:transp}
\end{equation}
Second, substituting this equality into \eqref{eq:ener}, we obtain
\begin{equation}
 \nu \int_{-\infty}^0 \left( |\MF{A}_+ (y)|^2 + |\MF{A}_- (y)|^2 \right) \D{} y =
 {\rm Im} \Bigl\{ \omega^2 \, \overline{\vec{\chi}}^\transp \bm{E} \vec{\chi} - g 
 \overline{\vec{\chi}}^\transp \bm{K} \vec{\chi} \Bigr\} .
\label{eq:ener'}
\end{equation}
In the same way as in \cite{KM1}, this yields the following assertion about the
kinetic and potential energy of the water motion.

\begin{proposition}\label{propos:1}
Let\/ $\bigl( \MF{\varphi} , \vec{\chi} \bigr)$ be a solution of problem\/
\eqref{eq:1}--\eqref{eq:5}, then
\begin{equation}
 \int_W |\nabla \varphi|^2\,\D{}x \, \D{}y < \infty \quad \mbox{and} \quad
 \nu \int_F |\varphi|^2 \, \D{}x < \infty .
\label{eq:finenerg}
\end{equation}
Moreover, the following equality holds:
\begin{equation}
 \int_W |\nabla \varphi|^2\,\D{}x \,\D{}y + \omega^2 \overline{\vec{\chi}}^\transp
 \bm{E} \vec{\chi} = \nu \int_F |\varphi|^2 \, \D{}x + g \,
 \overline{\vec{\chi}}^\transp \bm{K} \vec{\chi} .
\label{eq:lagrange}
\end{equation}
\end{proposition}

Here the kinetic energy of the water/body system stands in the left-hand side,
whereas we have the potential energy of this coupled motion in the right-hand side.
Thus the last formula generalises the energy equipartition equality valid when a
fixed body is immersed into water. Indeed, $\vec{\chi} = 0$ for such a body, and
\eqref{eq:lagrange} turns into the well-known equality (see, for example, formula
(4.99) in \cite{LWW}).

Proposition 1 shows that if $( \varphi, \vec{\chi} )$ is a solution of problem
\eqref{eq:1}--\eqref{eq:5} with complex-valued components, then its real and
imaginary parts separately satisfy this problem. This allows us to consider $(
\varphi, \vec{\chi} )$ as an element of the real product space $H^1 (W) \times
\RR^3$ in what follows (the sum of two quantities \eqref{eq:finenerg} defines an
equivalent norm in $H^1 (W)$).

\begin{definition}\rm
Let the subsidiary conditions concerning the equilibrium position (see \S~2) hold
for the freely floating body $\mhat{B}$. A non-trivial real solution $( \varphi,
\vec{\chi} ) \in H^1 (W) \times \RR^3$ of problem \eqref{eq:intid} and \eqref{eq:5}
is called a {\it mode trapped}\/ by this body, whereas the corresponding value of
$\omega$ is referred to as a {\it trapping frequency}.
\end{definition}

In order to determine when $( \varphi, \vec{\chi} ) \in H^1 (W) \times \RR^3$ is not
trapped by $\mhat{B}$ we write \eqref{eq:lagrange} as follows:
\begin{equation}
\vec{\chi}^\transp ( \omega^2 \bm{E} \vec{\chi} - g \, \bm{K} ) \vec{\chi}  = \nu
\int_F |\varphi|^2 \, \D{}x - \int_W |\nabla \varphi|^2\,\D{}x \D{}y .
\label{eq:eq}
\end{equation}
It is clear that the left-hand side is non-negative provided $\omega^2$ is
sufficiently large, and so we arrive at the following.

\begin{proposition}\label{propos:2}
Let $\bm{E}$ and $\bm{K}$ be given by \eqref{eq:EK} and let\/ $\omega^2$ be greater
than or equal to the largest $\lambda$ satisfying $\det( \lambda \bm{E} - g \bm{K} )
= 0$. If the domain $W$ is such that the inequality
\begin{equation}
\nu \int_F |\varphi|^2 \, \D{}x < \int_W |\nabla \varphi|^2\,\D{}x \D{}y
\label{eq:ineq}
\end{equation}
holds for every non-trivial $\varphi \in H^1 (W)$, then $\omega$ is not a trapping
frequency.
\end{proposition}

Note that if $W$ has finite depth, then $\nu$ must be changed to $k_0$ in relation
\eqref{eq:uas}, where the behaviour of the remainder must be also replaced by the
following one:
\begin{equation}
|r_\pm|, \ |\nabla r_\pm| = \MF{O} \bigl( |x|^{-1} \bigr) \ \mbox{as} \ |x| \to
\infty . \label{eq:fin_inf}
\end{equation}
In relations \eqref{eq:ener} and \eqref{eq:ener'} $\nu$ must be also changed to
$k_0$. On the other hand, formula \eqref{eq:transp} remains be valid in the same
form as above, and so proposition 2 is true in this case as well.

\subsection{Examples of water domains for which inequality \eqref{eq:ineq} holds}

We begin with the case when $W$ has infinite depth. By $\ell_d$ and $\ell_{-d}$ we
denote the rays emanating at the angle $\pi/4$ to the vertical from the points $(d,
0)$ and $(-d, 0)$, respectively, and going to the right and left, respectively.

Let the whole rays $\ell_d$ and $\ell_{-d}$ belong to $W$ for all $d > a$. Thus,
$B$ is confined within the angular domain between the lines inclined at $\pi/4$ to
the vertical and going through the points $(a, 0)$ and $(-a, 0)$ to the right and
left, respectively. Under this assumption, \cite{SU} proved (see also \cite{LWW},
\S\S\,3.2.2.1 and 3.2.2.2) that the inequality
\[ \nu \int_F |\varphi|^2 \, \D x \leq \int_{W_c} |\nabla \varphi|^2 \, \D x \D y 
\]
holds provided $\varphi$ satisfies conditions \eqref{eq:finenerg} and relations
\eqref{eq:1} and \eqref{eq:2}. Here $W_c$ is the subset of $W$ covered with rays
$\{ \ell_{d}:\ (d,0) \in F \} \cup \{ \ell_{-d}:\ (-d,0) \in F \}$. According to
the last inequality, if $\varphi$ is non-trivial, then \eqref{eq:ineq} holds.
Therefore, proposition~2 is applicable, thus giving a criterion which values of
$\omega$ are non-trapping frequencies for the freely floating $\mhat{B}$ whose
immersed part $B$ is confined as described above.

In order to obtain inequality \eqref{eq:ineq} in the case when $W$ has finite depth,
$\ell_d$ and $\ell_{-d}$ must be replaced by similar segments connecting $F$ and $H$
and inclined at a certain angle to the vertical that is a little bit less than
$\pi/4$. Numerical computations of \cite{SU} show that the same result as for deep
water is true when $B$ is confined between the segments inclined at
$44\frac{1}{3}^{\circ}$.

\section{Another criterion eliminating some particular \\ trapped modes}

In this section, we turn to the case when $B$ does not satisfy the conditions of
\S\,3.2. To be specific, we suppose that $W$ is bounded from below by the rigid
bottom $H$. Moreover, we assume that $\mhat{B}$ is symmetric about the $y$-axis (see
figure~\ref{fig1}); this implies that $N_1 = n_x$ $(N_2 = n_y)$ attains the opposite
(the same, respectively) values at every pair of points on $B$ which are symmetric
about the $y$-axis. Let also $\rho (x, y)$ be an even function of $x$, and so
$x^{(0)} = 0$ (the centre of mass lies on the $y$-axis); this implies that $N_3 = x
n_y - n_x (y-y^{(0)})$ has the same behaviour as $N_1$.

The last restriction on $\mhat{B}$ or, more precisely, on $B$ is expressed in terms
of the curves
\begin{equation}
x^2 + (y - a \cot \sigma)^2 = a^2 (\cot^2 \sigma + 1), \quad \pm x > 0 , \quad y < 0
, \label{eq:ca}
\end{equation}
parametrised by $\sigma \in (-\pi, 0)$. On curves of these two families we define
directions as shown in figure~\ref{fig1}. It is clear that all curves \eqref{eq:ca},
that intersect $H$ transversally, enter into $W$. Let this property also hold on
$S$; that is, all transversal intersections of curves \eqref{eq:ca} with $S$ are
points of entry into $W$ (see figure~\ref{fig1}). In what follows, a body satisfying
the listed conditions is referred to as belonging to the class ${\cal B}$ provided
the conditions considered in \S\,3.2 are not fulfilled for it.

The following assertion generalises the criterion of \cite{K} guaranteeing the
absence of trapped modes for fixed surface-piercing bodies immersed in deep water
and satisfying the above transversality condition with the family of curves
\eqref{eq:ca}. As in proposition~2 the values of $\omega$ that are not trapping
frequencies must be sufficiently large, but what is new that some restrictions must
be also imposed on the type of mode.

\begin{proposition}\label{propos:3}
Let\/ $W$ have finite depth and let\/ $\mhat{B}$ be a freely floating body belonging
to the class ${\cal B}$. If $\omega^2$ is strictly greater than the largest
$\lambda$ such that\/ $\det( \lambda \bm{E} - g \bm{K} ) = 0$ with $\bm{E}$ and
$\bm{K}$ given by \eqref{eq:EK}, then $\omega$ is not a trapping frequency for modes
of the form:

\vspace{1mm}

{\rm (a)} $\varphi$ is an even function of $x$ and $\vec{\chi} = (d_1, 0,
d_3)^\transp;$

{\rm (b)} $\varphi$ is an odd function of $x$ and $\vec{\chi} = (0, d_2, 0)^\transp$.
\end{proposition}

\vspace{-3mm}

\begin{proof}
Let us write relations \eqref{eq:1}--\eqref{eq:4} and \eqref{eq:H} using the bipolar
coordinates $(u, v)$. The corresponding conformal mapping is usually defined as
follows (see, for example, \cite{MF}, \S\,10.1):
\begin{equation}
x = a \sinh u / (\cosh u - \cos v) , \ \ \ y = a \sin v / (\cosh u - \cos v) .
\label{eq:cm}
\end{equation}
Therefore, \eqref{eq:cm} maps the strip $\{ -\infty < u < +\infty,\,-\pi < v < 0 \}$
onto $\RR^2_-$ so that for every $\sigma \in (-\pi , 0)$ the image of the left
(right) half-line $\{ \pm u > 0 , \, v=\sigma \}$ is the circular arc \eqref{eq:ca}
that lies in the left (right) half-plane (see figure~\ref{fig1}). Moreover, 
\[ \{ -\infty < u < +\infty,\,v=-\pi \} \quad \mbox{and} \quad \{ \pm u > 0,\,v=0 \}
\]
are mapped onto $\{ |x| < a,\,y=0 \}$ and $\{ \pm x > a,\,y=0 \}$ respectively.
Finally, we have that
\[ |z' (\zeta)| = a/(\cosh u - \cos v) , \quad \mbox{where} \ z = x + \ii y \ 
\mbox{and} \ \zeta = u + \ii v .
\]
The inverse mapping $\zeta (z)$ has the following properties: the points $a$ and
$-a$ on the $x$-axis go to infinity on the $\zeta$-plane, whereas $z = \infty$ goes
to $\zeta = 0$; thus $F$ is mapped onto the whole $u$-axis.

Denoting by ${\cal W}$ the image of $W$, we see that apart from the $u$-axis the
boundary $\partial {\cal W}$ includes the images of $S$ and $H$, say ${\cal S}$ and
${\cal H}$ respectively. According to properties of \eqref{eq:cm}, if $\mhat{B}$
belongs to the class ${\cal B}$, then ${\cal S}$ is symmetric about the $v$-axis,
lies within the strip $\{ -\infty < u < +\infty, -\pi < v <  - \alpha \}$ and
asymptotes the line $v = - \alpha$ as $u \to \pm \infty$; here $\alpha \in (0, \pi)$
is the angle between $S$ and $F$ at $(\pm a,0)$. Moreover, the right half of ${\cal
S}$ is the graph of a decreasing function of $u \in (0, +\infty)$; its maximum value
$v_b$ is the root of $\cos v - (a/b_0) \sin v = 1$. Finally, ${\cal H}$ is a closed
curve with the following properties. It is symmetric about the $v$-axis, is tangent
to the $u$-axis at the origin and is the graph of a concave function of $v \in (v_h,
0)$; here $v_h \in (-\pi, 0)$ is the root of $\cos v - (a/d) \sin v = 1$. It is
clear that $-\pi/4 < -\alpha < v_b < v_h < 0$.

Let $\phi (u,v) = \varphi (x(u,v),y(u,v))$, then relations
\eqref{eq:1}--\eqref{eq:4} yield that
\begin{equation}
\nabla^2 \phi = 0 \ {\rm in}\ {\cal W}, \ \ (\cosh u - 1) \phi_v = \nu a \, \phi \
{\rm when}\ v = 0, \ \ \nabla \phi \cdot {\bf n}_\zeta = \frac{\omega a {\bf
N}_\zeta^\transp \vec{\chi}}{\cosh u - \cos v} \ {\rm on}\ {\cal S} .
\label{eq:cbvp}
\end{equation} 
Here ${\bf n_\zeta}$ is the unit normal to ${\cal S} \cup {\cal H}$ exterior with
respect to ${\cal W}$ and ${\bf N}_\zeta = \vec{N}_{z (\zeta)}$. Moreover, condition
\eqref{eq:H} implies that
\begin{equation}
\nabla \phi \cdot {\bf n}_\zeta = 0 \ {\rm on}\ {\cal H} ,
\label{eq:calH}
\end{equation} 
whereas condition \eqref{eq:5} takes the form
\begin{equation}
\omega^2 \bm{E} \vec{\chi} = - \omega \int_{\cal S} \frac{\phi {\bf N}_\zeta \,
\D{}s_\zeta}{\cosh u - \cos v} + g \, \bm{K} \vec{\chi} . \label{eq:5'}
\end{equation}
Furthermore, conditions \eqref{eq:finenerg} give that
\begin{equation}
\int_{\cal W} |\nabla \phi|^2\,\D u \D v < \infty \ \ \mbox{and} \ \
\int_{-\infty}^{+\infty} \frac{\phi^2 (u,0)} {\cosh u - 1} \, \D u < \infty ,
\label{eq:ei2}
\end{equation}
whereas equality \eqref{eq:eq} turns into the following one:
\begin{equation}
\int_{{\cal W}} |\nabla \phi|^2 \, \D u \D v - \nu a \int_{-\infty}^{+\infty}
\frac{\phi^2 (u,0)}{\cosh u - 1} \, \D u  = - \vec{\chi}^\transp ( \omega^2 \bm{E}
- g \, \bm{K} ) \vec{\chi} .
\label{eq:eq'}
\end{equation}

Further considerations are based on the following identity (see \cite{LWW},
Subsection 2.2.2):
\begin{equation}
( 2u\phi_u + \phi ) \nabla^2 \phi = \nabla \cdot ( 2 u \phi_u + \phi ) \nabla \phi -
2 \phi^2_u - \left( u |\nabla \phi|^2 \right)_u .
\label{eq:VM}
\end{equation}
Here the left-hand side vanishes due to the Laplace equation for $\phi$. Let us
integrate this identity over ${\cal W}' = {\cal W} \cap \{ |u| < b \}$ and $b$ is
sufficiently large (in particular, ${\cal H} \subset \{ |u| < b \}$). Using the
divergence theorem, we get
\begin{eqnarray}
&& 2 \int_{{\cal W}'} \phi^2_u \, \D u \D v + \int_{{\cal S}' \cup {\cal H}} {\bf u}
\cdot {\bf n}\,|\nabla \phi|^2 \, \D S = \int_{-b}^{+b} \left[ 2 u \phi_u (u,0) +
\phi (u,0) \right]\,\varphi_v (u,0) \, \D u \nonumber \\ && \ \ \ \ \ \ \ \ \ \ \ \
\ \ \ \ \ \ \, + \int_{{\cal S}'} ( 2 u \phi_u + \phi) \, \nabla \phi \cdot {\bf
n}_\zeta \, \D S + \sum_{\pm} \pm \int_{\cal C_{\pm}} ( 2 u \phi_u + \phi)\,\phi_u
\, \D v,
\label{eq:ii1}
\end{eqnarray}
where ${\cal S}' = {\cal S}\cap \{ |u|<b \}$, ${\bf u} = (u,0)$, $\sum_{\pm}$
denotes the summation of two terms corresponding to the upper and lower signs,
respectively, and ${\cal C_{\pm}} = {\cal W}'\cap \{ u=\pm b \}$. All integrals on
the right arise from the first term on the right in \eqref{eq:VM} and one more
integral of the same type vanishes in view of the boundary condition \eqref{eq:calH}
on ${\cal H}$.

Let us consider each integral standing on the right in \eqref{eq:ii1}. Using the
free-surface boundary condition, we get that the first term is equal to
\begin{eqnarray*}
&& \nu a \int_{-b}^{+b} \left[ 2 u \phi_u (u,0) + \phi (u,0) \right] \, \frac{\phi
(u,0)}{\cosh u - 1} \, \D u \\ && \ \ \ \ = \nu a \int_{-b}^{+b} \frac{u \sinh u \,
\phi^2 (u,0)}{(\cosh u - 1)^2} \, \D u + \nu a \left[ \frac{u\,\phi^2 (u,0)}{\cosh
u - 1} \right]_{u=-b}^{u=+b} ,
\end{eqnarray*}
where the last expression is obtained by integration by parts. It follows from
\eqref{eq:finenerg} that $\varphi (x,y)$ tends to constants as $(x,y)\to (\pm a,0)$,
and so $\phi (u,v)$ has the same property as $u \to \pm \infty$. Therefore, the
integrated term in the last equality tends to zero as $b \to \infty$, whereas the
integral on the right converges in view of \eqref{eq:ei2}.

The second integral on the right in \eqref{eq:ii1} is equal to
\[ \omega a \int_{{\cal S}'} \frac{( 2 u \phi_u + \phi) \, {\bf N}_\zeta^\transp
\vec{\chi}}{\cosh u - \cos v} \, \D S .
\]
Since $S$ belongs to the class ${\cal B}$, we have that $\phi$ and $\varphi$ are
simultaneously even and odd functions of $x$ and $u$ respectively. Therefore,
either of the assumptions (a) and (b) implies that this integral vanishes because
the integrand attains opposite values at points of ${\cal S}'$ that are symmetric
about the $v$-axis.

Finally, \eqref{eq:ei2} implies that there exists a sequence $\{ b_k
\}_{k=1}^\infty$ tending to the positive infinity and such that the last sum in
\eqref{eq:ii1} tends to zero as $b_k \to \infty$. Passing to the limit as $k \to
\infty$, we see that the transformed equation \eqref{eq:ii1} with $b=b_k$ gives the
following integral identity:
\[ 2 \int_{\cal W} \phi^2_u \, \D u \D v + \int_{\cal S \cup {\cal H}}
{\bf u} \cdot {\bf n} \, |\nabla \phi|^2 \, \D S - \nu a \int_{-\infty}^{+\infty}
\frac{u \sinh u}{(\cosh u - 1)^2} \, \phi^2 (u,0) \, \D u = 0
\]
provided either of the assumptions (a) and (b) holds. 

Subtracting this from \eqref{eq:eq'} multiplied by two, we get
\begin{eqnarray}
&& \ \ \ \ \ \ 2 \int_{\cal W} \phi^2_v \, \D u \D v - \int_{{\cal S} \cup {\cal H}}
{\bf u} \cdot {\bf n}\,|\nabla \phi|^2 \, \D S \nonumber \\ && + \nu a
\int_{-\infty}^{+\infty} \frac{u \sinh u - 2(\cosh u - 1)}{(\cosh u - 1)^2} \,
\phi^2 (u,0) \, \D u \nonumber \\ && \ \ \ \ \ \ \ \ \ \ \ \ \ = - 2
\vec{\chi}^\transp ( \omega^2 \bm{E} - g \, \bm{K} ) \vec{\chi} \, .
\label{eq:final}
\end{eqnarray}
If $\omega^2$ is strictly greater than the largest $\lambda$ such that $\det(
\lambda \bm{E} - g \bm{K} ) = 0$, then cannot hold unless $\omega$ is not a trapping
frequency for modes of the form (a) and (b). Indeed, the right-hand side is negative
for such a value of $\omega$ and a non-trivial $\vec{\chi}$, whereas the left-hand
side is non-negative because $S$ belongs to the class ${\cal B}$ and the fraction in
the last integral is non-negative. The obtained contradiction proves the
proposition.
\end{proof}

\section{Conjecture}

Given the proof of a theorem guaranteeing the uniqueness of a solution to the
linearised problem about time-harmonic water waves in the presence of a fixed
obstacle, then this proof admits amendments transforming it into the proof of an
analogous theorem for the same obstacle floating freely with additional restrictions
on the non-trapping frequencies (they must be sufficiently large) and, in some
cases, on body's geometry and on the type of non-trapping modes.

\end{document}